\documentstyle[twoside,fleqn,espcrc2]{article}

\newcommand{\be}{\begin{equation}}
\newcommand{\ee}{\end{equation}}
\newcommand{\ba}{\begin{eqnarray}}
\newcommand{\ea}{\end{eqnarray}}
\newcommand{\AmS}{{\protect\the\textfont2
  A\kern-.1667em\lower.5ex\hbox{M}\kern-.125emS}}

\title{The quenched generating functional for hadronic weak interactions} 

\author{     E. Pallante\address{Institut f\"ur Theoretische Physik, 
Universit\"at Bern, Sidlerstrasse 5,\\
CH--3012 Bern, Switzerland }%
\thanks{supported by Schweizerisches Nationalfonds} }
       
\begin{document}

\begin{abstract}
The ultraviolet behaviour of the generating functional for hadronic weak
interactions with  $\vert\Delta S\vert =1,\, 2$ is investigated to one loop
for a generic number of flavours and in the quenched approximation.
New quenched chiral logarithms generated by the 
 weak interactions can be accounted for via a redefinition of the weak 
mass term in the $\Delta S=\pm 1$ weak effective Lagrangian at leading order.
Finally, we illustrate how chiral logarithms are modified 
by the quenched approximation in $K\to\pi\pi$ matrix elements with
$\Delta I=1/2$ and $3/2$.
\end{abstract}

\maketitle

\section{Introduction}
The computation of weak matrix elements at long distances is one of the main
and still open problems in lattice QCD. Much progress has been done, although
their computation still suffers  from the presence of major
sources of systematic errors: finite volume effects, unphysical quark masses,
 the need of computing {\em unphysical} matrix elements \cite{BERNARD}
 (see also \cite{MARTI} for recent alternative proposals) and 
finally the fact that still most of the lattice evaluations are done in the
quenched approximation. Here we report on some results concerning formal
aspects of the quenched approximation in the ultraviolet regime. The framework
we use is known as Chiral Perturbation Theory (ChPT) \cite{gl84} 
 and quenched ChPT \cite{qCHPT,npb}, extended to full (unquenched) 
 weak interactions in \cite{KMW} and to the quenched case in \cite{EP1}.
The advantages of deriving the generating functional of the effective low
energy theory in the full and quenched case are at least two: first, it allows
for a systematic control on the quenched modifications and second, it gives in
one step the coefficients of all chiral logarithms for any Green's function or
S-matrix element.

\section{The weak generating functional to one loop}

The quenched generalization of the weak effective Lagrangian for nonleptonic
 weak
interactions with $\vert\Delta S\vert =1,\, 2$ and
 at leading order in the chiral
expansion can be written as follows
\ba
{\cal L}_{{\tiny{\Delta S=1}}}&=&
\tilde{V}_8(\Phi_0) \mbox{str}(\Delta_{s 32}u_{\mu s}u^\mu_s )\nonumber\\
&&\hspace{-1.7cm}+\tilde{V}_5(\Phi_0)\mbox{str}(\Delta_{s 32}\chi_{+s} )
\nonumber\\
&&\hspace{-1.7cm}
+\tilde{V}_0(\Phi_0)\mbox{str}(\Delta_{s 32}u_{\mu s}) \mbox{str}(u^\mu_s )
+ {\mbox{h.c}} \nonumber\\
&&\hspace{-1.7cm}+
\tilde{V}_{27}(\Phi_0)t^{ij,kl}\mbox{str}(\Delta_{s ij}u_{\mu s})
\mbox{str}(\Delta_{s kl}u^\mu_s) 
\label{WEAK}
\ea
and the same last term for $\Delta S=2$ with the appropriate $t^{ij,kl}$
tensor. 
The projection matrix onto the octet and 27-plet 
components is the graded matrix 
$ \Delta_{s ij}=u_s \lambda_{ij} {1+\tau_3\over 2}u_s^\dagger$, with
$\left (\lambda_{ij}\right )_{ab} = \delta_{ia}\delta_{jb}$.
The graded fields
$u_{\mu s}=iu_s^\dagger D_\mu U_s u_s^\dagger = u_{\mu s}^\dagger$ and 
$\chi_{+s}=u_s^\dagger\chi_s u_s^\dagger +u_s\chi_s^\dagger u_s$ contain the
dynamical meson field $U_s\equiv u_s^2 = \exp (\sqrt{2} i\, \Phi/F)$ 
in the usual way.
The field $\chi_s = 2B_0{\cal M}_s +\ldots$ contains the graded
quark mass matrix.
The potentials $\tilde{V}_i(\Phi_0)$ are real and even
functions of the super-$\eta^\prime$ field $\Phi_0 = \mbox{str}(\Phi)$ and we
keep them up to order $\Phi_0^2$.
The ultraviolet divergences of the 
weak generating functional to one loop can be derived as it was done 
in the strong sector \cite{npb}, provided an expansion in powers of $G_F$ is
performed.  For degenerate quark masses we get for the $\Delta S=\pm 1$ octet
contributions at order $G_F$:
\begin{eqnarray}
Z_{(8)}^q&=&
-\frac{1}{(4\pi)^2(d\!-\!4)} \int \! dx\, \left\{
{1\over 6}m_0^2(g_8^\prime
  -2g_8) k \langle\Delta\chi_+\rangle \right.  \nonumber\\
&&\hspace{-1.7cm} +{1\over 4}g_8 W_4  
 +{1\over 8}W_6
+\left ({\alpha\over 12}(g_8-g_8^\prime )-{1\over 8}\bar{g}_8\right ) 
\left ( W_{12}+W_{36}\right ) \nonumber\\ 
&&\hspace{-1.7cm} +\left [-{3\over 16}g_8 
\left ( 1-{16\over 3}v_1\right ) +{1\over
    8}g_8^\prime \left (1-4v_1\right ) \right ] W_7  \nonumber\\ 
&&\hspace{-1.7cm} +{1\over 16}(g_8-8\tilde{v}_8)W_8  
 +\left ({\alpha\over 12}(2g_8-g_8^\prime )-{1\over 8}\bar{g}_8\right )W_{10} 
\nonumber\\
&&\hspace{-1.7cm}
+ \left [ \left ({1\over 8}+{\alpha^2\over 36}\right )(g_8^\prime-g_8)
  -{1\over 2} v_2 (g_8^\prime -2g_8) -{1\over 2} \tilde{v}_5 \right.
\nonumber\\
&&\left.\left.  \hspace{-1.7cm} +{\alpha\over
    12}\bar{g}_8 \right ] W_{11} 
  + \left (-{\alpha\over 6}g_8+{1\over 4}\bar{g}_8\right ) 
(W_{21}+W_{22})  \right\}.\nonumber
\end{eqnarray}
The operators $W_i$ are listed in Appendix A of \cite{EP1}.
The 27-plet contribution is given by
\ba
Z_{(27)}^q&=&
-\frac{1}{(4\pi)^2(d\!-\!4)} \int \! dx\, g_{27}\left\{ 
{1\over 12}D_1+{5\over 6}D_2\right.\nonumber\\
&&\hspace{-1.7cm}-{1\over 8}D_3-{7\over 24}D_4-{1\over 4}D_5
+{1\over 4}D_6+{1\over 2}D_7+{3\over 8}D_8\nonumber\\
&&\hspace{-1.7cm}
+{1\over 4}\left (
  1-2{\tilde{v}_{27}\over g_{27}}\right ) D_{10}
-{3\over 8}D_9-{1\over 8}\left (1-{2\over 3}\alpha\right ) D_{12}
\nonumber\\
&&\hspace{-1.7cm}
-{1\over 4}D_{11} -{1\over 12}D_{13} +{1\over 12}D_{14}-{1\over 24}D_{15}
-{1\over 12}D_{16}\nonumber\\
&&\hspace{-1.7cm}+ {7\over 12}D_{17} +  {7\over 12}D_{18} -{1\over 4}D_{19} -{1\over 4}D_{20}
+{1\over 6}D_{21}\nonumber\\
&&\hspace{-1.7cm}\left. +{1\over 6}D_{22}+{1\over 4}D_{23}+{1\over 4}D_{24} 
\right\}  +O(G_F^2)\, ,\nonumber
\ea
where the operators $D_i$ are listed in Appendix A of \cite{EP1}.
$m_0^2$ and $\alpha$ are the usual singlet parameters.
The couplings $g_8,\, g_8^\prime,\, \bar{g}_8$ and $g_{27}$ are the first
terms in the expansion of the weak potentials $\tilde{V}_i,\, i=8,5,0,27$,
while the parameters $\tilde{v}_i$ are the coefficients of the $\Phi_0^2$ term.
The quenched generating functional for $\Delta S=2$ interactions has 
exactly the same
structure of $Z_{(27)}^q$, with the appropriate $t^{ij,kl}$ tensor.

The quenched approximation largely reduces the ultraviolet
 divergent contribution to
the octet sector at one loop. Of the initially divergent 25 octet operators
 only 10 remain in the quenched approximation.
The octet operators $W_5,\,\ldots W_{12}$ and $W_{38}$  contribute 
to $K\to 2\pi$ decays.
In the 27-plet sector all the unquenched chiral invariants $D_i$ survive to
the quenched approximation.
Only the operators $D_8,\,\ldots D_{12}$ contribute to 
$K\to 2\pi$ decays.

The systematic cancellation of the flavour number dependence follows the rules
outlined in \cite{npb,plb}. The 
cancellation of the $1/N, 1/N^2$ terms is
provided by the sum of the non-singlet contributions and the singlet 
contributions within the bosonic sector. This type of cancellation is entirely
due to the presence of a dynamical singlet field.
The linear flavour number dependence is cancelled by the fermionic ghost
determinant as expected.

\subsection{ Quenched chiral logarithms}

The most relevant behaviour of the quenched generating functional to
one loop is the appearance of quenched chiral logarithms, i.e. of the
type $m_0^2\log m_\pi^2$, that are pure artefacts of the quenched
approximation. As it was discussed at length in \cite{npb}, quenched
chiral logarithms appearing in the strong sector can be formally
reabsorbed in a redefinition of the $B_0$ parameter.  In the weak
sector an analogous mechanism occurs. The quenched chiral logarithms
which appear through the first term in the equation for $Z_{(8)}^q$
can be formally reabsorbed into a redefinition of the weak mass term
coupling $g_8^\prime$ of the leading order Lagrangian (\ref{WEAK}). To
remove the $m_0^2$ divergence one has to add to the lowest order
parameter $g_8^\prime$ a d-dependent part proportional to $m_0^2$ that
has a pole at d=4 and the rescaled coupling can be defined as follows:
\ba g_{8R}^\prime&=&\nonumber\\ &&\hspace{-1.5cm} g_8^\prime\left (
1-{m_0^2\over 48\pi^2F^2}\left (1-2{g_8\over g_8^\prime}\right )\log
{M^2\over \mu^2} +\delta g_8^\prime (\mu )\right )\, .\nonumber
%\label{QLOGS}
\ea   
The rescaling of the coupling $g_8^\prime$ together with the rescaling of
the parameter $B_0\to\bar{B}_0$ defined in \cite{npb} 
in the tree level contribution to
any weak observable can be used as a short-cut procedure to
unreveal the presence of quenched chiral logarithms, generated 
when the quenched approximation is implemented to one loop.

\section{$K\to \pi\pi$ matrix elements }

 Using the one loop expression of  $K\to \pi\pi$ matrix elements
 as derived in \cite{BPP} and the quenched counterpart
of the ultraviolet divergences derived here, we can produce
a few quantitative estimates of quenching effects on the coefficients of  
chiral logarithms that contribute to $K\to \pi\pi$ amplitudes. We work in the
infinite volume limit for illustrative purpose. In \cite{E2} we shall report
on the analysis of {\em unphysical} choices of the kinematics used for the
computation of lattice $K\to \pi\pi$  matrix elements. 

We decompose the $K\to \pi\pi$ matrix elements into 
definite isospin invariant amplitudes:
$\Im m A_0\equiv \Im m (A_0^8+ A_0^{27})$ with $\Delta I=1/2$, and $\Im m A_2$
with $\Delta I=3/2$.
%%%%%%%%%%%%%%%%%%%%%%%%%%%%%%%%%%%
\begin{table*}[hbt]
\setlength{\tabcolsep}{1.5pc}
\newlength{\digitwidth} \settowidth{\digitwidth}{\rm 0}
\catcode`?=\active \def?{\kern\digitwidth}
% -----------------------------------------------------
\caption{ The coefficients of the one loop chiral logarithms in $K\to\pi\pi$
  amplitudes, full and quenched.}
\label{QFULL}
%\begin{center}
\begin{tabular*}{\textwidth}{@{}l@{\extracolsep{\fill}}rrrr}
\hline
 & Full ($m_K=m_\pi$) &  Full ($m_\pi =0$) & Quenched \\
\hline
 &&&\\
  $A_0^8$   & ${11\over 18}$ & $-{5\over 4}$  & $-\left ({1\over 2}+{4\over
 9}\alpha^2 -2\alpha\right )+v_i,\tilde{v}_i,\bar{g}_8,g_8^\prime$ \\
 &&&\\
  $A_0^{27}$   & $-{17\over 4}$ & $-{15\over 2}$  & $-4
\left (1-{\alpha\over 2}\right )$ \\
 &&&\\
  $A_2$   & $-{11\over 2}$ & $-{3\over 4}$  &$-{13\over 4}$ \\
 &&&\\
\hline
\end{tabular*}
%\end{center}
\end{table*}
In Table (\ref{QFULL}) we compare the coefficients of chiral logarithms for
the three amplitudes in the full theory and in the quenched approximation.
For the full amplitudes we consider two extreme mass configurations in the one
loop corrections: a) degenerate masses, i.e. $m_K=m_\pi$ and b) $m_\pi =0$.
The numerical analysis of the physical non degenerate mass case together with
the comparison with {\em unphysical} choices of the matrix elements on the
lattice will be given in \cite{E2}.
The coefficients of chiral logarithms in the full 27-plet amplitudes ${\Im}m
 A_0^{27}$ and  ${\Im}m A_2$ are quite large in the degenerate mass
  limit. In addition, by the comparison of second and third column in 
Table (\ref{QFULL})
one can conclude that all the coefficients in the full amplitudes are
  extremely sensitive to the variation of masses. 
Using $\alpha\simeq 0.6$ and disregarding for now the unknown contributions to
the octet amplitude in the quenched case we find the following pattern going 
from the full degenerate mass case  to the quenched one.
Quenching reduces by a tiny amount (from 0.6 to 0.54) the coefficient of
the chiral logarithm in the octet amplitude $A_0^8$, while 
reduces the one in $A_{27}$ by about $34\%$ in absolute value and  $41\%$ 
in the case of $A_2$.
Note that the pattern is opposite for $A_2$ when we compare the quenched
amplitude to the $m_\pi =0$ limit of the full amplitude. 
This 
analysis (see \cite{EP1,E2} for more details) 
shows that quenching largely affects the coefficient of the chiral
logarithm in the 27-plet $\Delta I=3/2$ and $\Delta I=1/2$ amplitudes. In
addition, the comparison with the $m_\pi =0$ limit of the full amplitudes
(expected to be the most approximate to the physical value)
shows that the modification induced by quenching follows
a pattern that tends to suppress the $\Delta I=1/2$ dominance.

%%%%%%%%%%%%%%%%%%%%%%%%%%%%%%%%%%%%%%%

\end{document}